\documentstyle[epsf,twocolumn]{mn}
\def\centerline#1{\hbox to\textwidth{\hss#1\hss}}
\def\spose#1{\hbox to 0pt{#1\hss}}
\def\simlt{\mathrel{\spose{\lower 3pt\hbox{$\mathchar"218$}}
     \raise 2.0pt\hbox{$\mathchar"13C$}}}
\def\simgt{\mathrel{\spose{\lower 3pt\hbox{$\mathchar"218$}}
     \raise 2.0pt\hbox{$\mathchar"13E$}}}
\def\simpropto{\mathrel{\spose{\lower 3pt\hbox{$\mathchar"218$}}
     \raise 2.0pt\hbox{$\propto$}}}

\title{RXTE highlights of 34.85-day cycle of Her X-1}
\author[N.I.Shakura et al.]
{N.I.Shakura$^1$, N.A.Ketsaris$^1$, M.E.Prokhorov$^1$ and K.A.Postnov$^2$\\
$^1$ Sternberg Astronomical Institute,
119899 Moscow, Russia\\
$^2$ Faculty of Physics, Moscow State University, 119899 Moscow, Russia}

\date{Accepted 1998...
      Received 1998 ...}
\pagerange{\pageref{firstpage}--\pageref{lastpage}}
\pubyear{1998}

\begin{document}
\maketitle

\label{firstpage}

\begin{abstract}

An analysis of the publically available RXTE archive on Her X-1 including
data on 23 34.85-day cycles is performed. The turn-on times for these cycles
are determined. The number of cycles with a duration of 20.5 orbits has been
found exceedingly larger than of shorter (20 orbits) or longer (21 orbits).
A correlation between the duration of a cycle and its mean X-ray flux is
noted. The mean X-ray light curve shows a very distinct short-on state. The
anomalous X-ray absorption dip is found during the first orbit after the
turn-on in the main-on state for the cycles starting near the binary phase
0.25, while is present during two successive orbits in the low-on state. The
post-eclipse recovery feature have not been found in the main-on state but
appears at least for two orbits during the low-on state. The pre-eclipse
dips are present both in main-on and low-on state and demonstrate the
behaviour as in early observations. The comparison of durations of the
main-on and short-on states enabled us to constrain the accretion disk
semi-thickness and its inclination to the orbital plane.

\end{abstract}

\section{Introduction}

Her X-1 is one the first X-ray binary pulsar discovered
by the eminent UHURU
satellite in 1972 (Tananbaum et al. 1972, Giacconi et al. 1973) and since
then remains one of the most studied accretion-driven X-ray stellar
binaries. The pulsar is an accretion-powered magnetized rotating ($P=1.24 $s~)
neutron entering an eclipsing X-ray binary with a low-mass ($\sim 2 M_\odot$)
optical companion in a practically circular orbit with a period
of 1.7 d. The binary orbit inclination is $i=85-88^o$, the total
duration of the X-ray eclipse is around 20,000 s.
The optical counterpart, HZ Her, first
suggested by Liller (1972),  provides a classical example of
the reflection effect when the strong orbital modulation in optics is due to
a powerful illumination of the part of normal star atmosphere facing
X-ray source (Bahcall \& Bahcall 1972; Cherepashchuk et al. 1972).
The amplitude of the reflection effect strongly increases
in ultraviolet and according to many optical UBV-photometrical data is
$\Delta m_V =1.45,
\Delta m_B =1.6, \Delta m_U=2.55 $.

Already very first UHURU observations of Her X-1 revealed the presence of
a long-term 34.85-d cycle. Its properties have been so interesting
that practically all specialized X-ray satellites studied it
(among which
Copernicus ,
Ariel-5, Ariel-6 (Davison \& Fabian 1974, 1977, Ricketts et al. 1982),
HEAO-1 (Gorecki et al. 1982, Soong et al. 1987),
Hakucho  and TENMA (Nagase et al. 1984, Ohashi et al. 1984),
EXOSAT (\"Ogelman \& Tr\"umper 1988),
Ginga (Deeter et al. 1991),
Astron (Sheffer et al. 1992),
BATSE (Wilson et al. 1994),
and RXTE (see
http://space.mit.edu/XTE/asmlc/srcs/data/herx1\_105.dat)).
RXTE data are shown in Fig. 1.

The gross shape of the X-ray light curve consists of a main-on X-ray state
with a mean duration of 7 orbital periods surrounded by two off-states each 4
orbital cycles in duration, and of a secondary short-on state of smaller
intensity with a typical duration of $\sim 5.5$ orbital cycles. The form of
the X-ray light curve is asymmetric: in the main-on state the X-ray
intensity rapidly increases during 1.5-2 orbital cycles, stays at maximum
over 2.5 cycles, and then decreases to a minimum in the successive 3 orbital
revolutions. At the beginning of the ascending branch a strong low-energetic
absorption dominates in the X-ray spectrum while at the descending part it
does not vary appreciably. A characteristic feature of the X-ray light curve
is the presence of strong pre-eclipse and anomalous absorption dips during
the main-on state. The anomalous dips are generally found only in the
first or second orbit after turn-on near orbital phase 0.55. The pre-eclipse
dips march from the eclipse toward earlier orbital phase in successive
orbits. Sometimes, a post-eclipse recovery is observed in the second
orbit after turn-on.

The short-on X-ray state was found with the Copernicus 
satellite (Fabian et al. 1973) and later observed with Ariel-5 
(Cooke \& Page 1975). Subsequently, Jones \& Forman
(1976) found the short-on state in the UHURU records.

Most observed details of the X-ray light curve can be naturally explained by
the model of tilted accretion disk undergoing the counter-precession
opposite to the orbital motion (Gerend \& Boynton 1976, Crosa
\& Boynton 1980).
At the main-on state, the disk is mostly opened to the observer (the total
inclination angle represents a combination of the binary and disk
inclinations), at the low-on state the opening angle is smaller.
The X-ray source turn-off seen for a few orbits is due to the screening
by the thick accretion disk. In fact, a weak X-ray glow is observed during
the off-states, which is apparently due to scattering on the electrons of an
extended accretion disk corona (Jones \& Forman 1976).

\begin{figure*}
\epsfxsize=\textwidth
\fbox{\vbox{
\centerline{}
\centerline{This figure is avaliable at URLs:}
\centerline{}
\centerline{\bf http://xray.sai.msu.su/pub/preprints/Prokhorov/rxte/fig1.ps}
\centerline{or}
\centerline{\bf http://xray.sai.msu.ru/pub/preprints/Prokhorov/rxte/fig1.ps}
\centerline{}
}}
\caption{RXTE light curve of Her X-1 in counts per second as a function
of MJD. Error bars (in the bottom right corner) show minimal, mean, and
maximal 1-$\sigma$ errors of individual points. }
\end{figure*}

As was discovered by UHURU and confirmed by the
subsequent observations,
the main
turn-on times preferentially occur around 0.25 and 0.75 orbital phases of
the binary system. Levine \& Jernigan (1982)
suggested that this tendency is due to nutation effects of an inclined
precessing accretion disk.

Staubert et al. (1983) analysed turn-on data and
concluded that
the actual separation between two successive turn-ons may be
either 20.0, 20.5 or 21.0 binary cycles selected random
with equal statistical probabilities. 
Baykal et al. (1993) revealed that the statistical interpretation of
turn-on behaviour is consistent with a white-noise process in the first
derivative of the 35-d phase fluctuations (or a random walk in 
clock phase).
It remains unclear up to now
why the precessional period is so close to 20.5 orbital
cycles ($1.7\times 20.5 = 34.85$ days !).

The traces of 34.85-d cycle have also been found in optical
modulation of HZ Her (Kurochkin 1972).
Gerend \& Boynton (1976) modeled
the optical light curve by thick precessing tilted
accretion disk. Afterwards.
Howarth \& Wilson (1983), using more extended  photometrical data,
came to similar conclusions.

Variability of the source on longer time scales is also
very surprising.
In 1983, the EXOSAT observations
detected a 9-months' turn-off of the
X-ray source (IAUC 3852, 1983) during which the reflection effect
was however persistent in optics (Mironov et al. 1984).
Notably, the subsequent normal turn-on of the X-ray source occurred
practically in phase with the mean ephemeris calculated
for 20.5 orbital cycles periodicity (\"Ogelman et al. 1985) and since then the source
have not shown visible deviations from the mean schedule.

A few episodes of significant decrease of the optical
reflection effect and fading of HZ Her
have been found in Zonnenberg
archive optical plates (Wentzel \& Gessner 1972).
Jones et al. (1973) analysed almost 100-years' Harward phototeque
and found also several episodes of the absence of the reflection effect
with a duration from ten days to years. During such off-states
the light curve of  HZ Her decreased in amplitude down to $\sim 0.^m3$,
had a  double-wave shape, and was solely due to  ellipsoidal form
of HZ Her.

In spite of the wealth of X-ray and optical data,
the nature of the 34.85-d cycle in Her X-1 still remains
controversial. In addition to the model of precessing
accretion disk, a free precessing neutron star was
suggested to underly 34.85-d clock mechanism
(Brecher 1972, Shklovskii 1973, Novikov 1973, etc.).
The evidence of the free precessing neutron star
was found in EXOSAT observations (Tr\"umper et al. 1986).
Recently, to explain the rapid change of X-ray pulse form
during the main-on state observed by HEAO-1 in 1978, we
suggested a model of free precessing triaxial neutron star
(Shakura et al. 1998).

With the launch of all-sky X-ray monitors (ASM) onboard
Compton Gamma-Ray Observatory (CGRO) (experiment
BATSE) and Rossi X-ray Time Explorer (RXTE) satellites,
the new possibility emerged
of collecting continuous X-ray data and thereby
of studying long-term X-ray behaviour of Her X-1.
This advantage of BATSE observations has been used by
Wilson et. al (1994) to reveal a significant (but not
universal, see a detailed review by
Bildsten et al. 1997)  correlation between
the X-ray pulsar frequency derivative $\dot \nu$ and X-ray flux
at the peak of the main-on portion of 34.85-d cycle, and to
confirm a possible
correlation of early turn-ons with
decreased mass transfer rate suggested by \"Ogelman (1987).

The purpose of this paper is to analyse properties of 34.85-d
X-ray cycle of Her X-1  using publically available
RXTE data.

{
\def\e{}
\def\c#1{#1}
\def\l#1{\multicolumn{1}{@{}l@{}}{#1}}
\def\C#1{\multicolumn{2}{c}{#1}}
\tabcolsep=1mm
\begin{table*}
\begin{center}
\caption{Turn-on time determination data}
\begin{tabular}{rcrr@{}lr*{3}{ccr}}
\hline\hline
No & state & \multicolumn{1}{c}{$MJD$}& \C{$\Phi_b$}& \multicolumn{1}{c}{$\Delta$}
&& \multicolumn{2}{c}{Tmpl[$\Phi_b]$} &&
\multicolumn{2}{c}{Tmpl[$\Phi_b-0.5$]} &&
\multicolumn{2}{c}{Tmpl[$\Phi_b+0.5$]} \\
\cline{8-9}
\cline{11-12}
\cline{14-15}
    &   &           &         &       &     && $t/u$    & $p$(\%) && $t/u$   & $p$(\%)&&
$t/u$  & $p$(\%)\\
\hline
253 & m & 50146.565 & $>$0.07 &$-$0.24&     &&   10/7 & 95  &&   9/6 &  13&&  17/13&  42 \\
    &   &           &         &       & 20.5&&   18/11& 93  &&\\
254 & m & 50181.419 & 0.65    &$-$0.73&     &&    7/5 & 90  &&  12/8 &  13&&  \e   &   0 \\
    &   &           &         &       & 20.5&&   18/14& 86  &&\\
255 & m & 50216.272 & 0.22    &$-$0.34&     &&   10/7 & 78  &&  \e   &   0&&  18/14&  95 \\
    &   &           &         &       & 20.5&&   \e   &     &&\\
256 & m & 50251.126 &\C{$\sim$0.75}   &     &&   18/15& 78  &&  \e   &   0&&  \e   &   0 \\
    &   &           &         &       & 20.5&&   \e   &     &&\\
257 & m & 50285.979 & 0.21    &$-$0.24&     &&   10/7 & 77  &&  \e   &   0&&  10/7 &  18 \\
    &   &           &         &       & 20.5&&   \e   &     &&\\
258 & m & 50320.833 &\C{$\sim$0.75}   &     &&    9/6 & 99  &&  10/7 &  11&&  \e   &   0 \\
    &   &           &         &       & 20.5&&   \e   &     &&\\
259 & m & 50355.686 &\C{$\sim$0.25}   &     &&    6/5 & 52  &&   7/5 &   7&&  \e   &   0 \\
    &   &           &         &       & 20.5&&   18/10& 52  &&\\
260 & m & 50390.540 & 0.69    &$-$0.72&     &&   13/10& 83  &&  \e   &   0&&  \e   &   0 \\
    &   &           &         &       & 20.5&&   \e   &     &&\\
261 & h & 50425.393 & 0.21    &$-$0.25&     &&   12/9 & 99  &&  \e   &   0&&  \e   &   0 \\
    &   &           &         &       & 20.5&&   \e   &     &&\\
262 & h & 50460.246 &\C{$\sim$0.75}   &     &&   18/13& 99  &&  \e   &   0&&  10/8 &  24 \\
    &   &           &         &       & 21.0&&   \e   &     &&\\
263 & l & 50495.950 & 0.61    &$-$0.80&     &&   14/9 & 92  &&  \e   &   0&&  \e   &   0 \\
    &   &           &         &       & 20.5&&   \e   &     &&\\
264 & m & 50530.803 & 0.20    &$-$0.40&     &&    6/5 & 91  &&  \e   &   0&&  12/9 &  49 \\
    &   &           &         &       & 20.5&&   18/11& 97  &&\\
265 & h & 50565.657 & 0.62    &$-$0.70&     &&   18/13& 81  &&  \e   &   0&&  \e   &   0 \\
    &   &           &         &       & 20.5&&   \e   &     &&\\
266 & h & 50600.510 &\C{$\sim$0.25}   &     &&   12/6 &$>$99&&  11/6 &  87&&  18/10&   5 \\
    &   &           &         &       & 21.0&&   \e   &     &&\\
267 & m & 50636.214 &\C{$\sim$0.25}   &     &&    6/5 & 70  &&  \e   &   0&&  \e   &   0 \\
    &   &           &         &       & 20.0&&   18/11& 70  &&\\
268 & m & 50670.217 &\C{$\sim$0.25}   &     &&   16/5 & 71  &&  \e   &   0&&  \e   &   0 \\
    &   &           &         &       & 20.5&&   \e   &     &&\\
269 & h & 50705.071 & 0.63    &$-$0.67&     &&    6/5 & 88  &&  \e   &   0&&  \e   &   0 \\
    &   &           &         &       & 20.5&&    9/7 & 84  &&\\
270 & l & 50739.924 &\C{$\sim$0.25}   &     &&    6/6 & 88  &&   6/6 &  34&&  \e   &   0 \\
    &   &           &         &       & 20.0&&   14/11& 85  &&\\
271 & l & 50773.927 & $>$0.07 &$-$0.31&     &&   18/8 & 76  &&  \e   &   0&&  \e   &   0 \\
    &   &           &         &       & 20.0&&   \e   &     &&\\
272 & l & 50807.931 &\C{$\sim$0.25}   &     &&   17/12& 72  &&   7/5 &  33&&  \e   &   0 \\
    &   &           &         &       & 20.5&&   \e   &     &&\\
273 & m & 50842.784 &$>$0.5   &$-$0.74&     &&    7/5 & 77  &&  \e   &   0&&  \e   &   0 \\
    &   &           &         &       & 20.5&&   18/14& 80  &&\\
274 & m & 50877.639 & 0.27    &$-$0.39&     &&    6/6 & 78  &&  \e   &   0&&  \e   &   0 \\
    &   &           &         &       & 20.5&&   10/8 & 77  &&\\
275 & l & 50912.492 & 0.63    &$-$0.71&     &&   13/10& 99  &&  \e   &   0&&  \e   &   0 \\
    &   &           &         &       &$>$20&&   18/15&$>$99&&\\
\hline\hline
\end{tabular}
\end{center}
\end{table*}
}

\section{The data and analysis}
\subsection{Turn-on time determination}

RXTE data archive contains X-ray (2-12 keV) count rates averaged over
predominantly 90-s time intervals (Fig.~1) started from MJD 50087. New data
is added daily to the archive. Despite some gaps in the data, the archive is
especially useful in reconstructing the mean X-ray light curve by means of
superposing many cycles with account of turn-on times phases. In addition,
it is possible to determine the turn-on times of 35-day cycle with a good
accuracy.

This accuracy is limited by the RXTE observations (the maximum UHURU
count rate from Her X-1 was about 100 counts per second while that of RXTE is
about $\le 10$ counts per second);, nevertheless, for 
cycles with no gaps in data, one may clearly distinguish 
around which orbital phase, 0.25 or 0.75, each cycle was turned-on 
(see Table 1).


To construct the mean X-ray light curve of Her X-1, we
first determined turn-on times. Using the ephemeris of X-ray
eclipse
$$
t_{min}=JD 2441329.57519+1^d.70016773\times E
$$
as given by Deeter et al. (1981)
\footnote{A slight decrease in orbital period of Her X-1 discovered by
Deeter et al. (1991) $\dot P_b/P_b=(-1.32\pm 0.16)\times 10^{-8}$
yr$^{-1}$ has no effect on our results.},
we reduced all data to the Solar system
barycenter.

For well-filled cycles, we have inspected the binary phase
intervals about $\sim 0.25$ and $\sim 0.75$ preceding the
X-ray source turn-on. For this purpose, we have combined 
consecutive data counts into groups such that the time
interval between each two adjacent points inside a group 
does not exceed $0.01$ JD. The group 
has been centered to the mean time of the points inside it.     
 

The observed
count rates, $C_i$, were averaged inside these groups
with a weight inversely proportional to
variance $\sigma_i^2$
$$
\langle  S \rangle = \frac{\sum\limits_{i=1}^{n}\frac{C_i}{\sigma_i^2}}
{\sum\limits_{i=1}^n\frac{1}{\sigma_i^2}}\,,
$$
where $n$ is the total number of individual points in each group.
After that we determined the total variance in each group as
$$
D^2=\frac{1}{n}\sum\limits_{i=1}^n\sigma_i^2+
\frac{n}{n-1}
\frac{\sum\limits_{i=1}^n\frac{\left(\langle S \rangle -
C_i\right)^2}{\sigma_i^2}}{\sum\limits_{i=1}^n\frac{1}{\sigma_i^2}}
$$

Next we have compared the signal-to-noise ratio 
$$
\frac{S}{N}=\langle S \rangle \frac{\sqrt{n}}{D}
$$ 
of the neighbour groups until 
a group with $S/N\simgt 3\sigma$ has been found. The turn-on time
has been considered to lie between this groups (see Table 1).
This methodics has enabled us to determine the  turn-on times
for 14 cycles 
with a typical accuracy of $\Delta\Phi_b\sim 0.05-0.08$.
Due to gaps in data, 
for 3 cycles (\#\# 259, 262, 270) the turn-on times have been
to a half-orbit accuracy, and for remaining 6 cycles
the data gaps are so large that the accuracy of the turn-on
time determination is more than one orbital period.  

To recover the turn-on times of 9 cycles with data gaps, 
using 14 well-defined cycles, we have constructed
templates, separately for cycles turned-on at $\Phi_b=0.25$ and 0.75, by
averaging all "0.25-cycles" and "0.75-cycles" inside 0.068-0.5 and 0.5-0.932
orbital phase intervals (excluding X-ray eclipses).
Of course, actual turn-on phase of a, say,
"0.25-cycle" differs from 0.25, but when averaging no phase shifts were made
for different cycles. 

Since the mean X-ray amplitude of the cycles varies, no unique template can
be constructed. So we have categorized all cycles into three X-ray intensity
groups (high, medium, low; see Table 1) and have constructed templates for
each group separately.
Cycles \# 263 and \# 275 can be classified as 
``conditionally low'' since they are morphologically different
from all other cycles,  high, low, and medium: 
their intensity reaches maximum 
during 3 orbits while all others during 1-1.5 orbital periods. 

After that we have checked each template to the data-poor cycles,
with the cycle data having been preliminary averaged in the 
same phase bins as the template points.
Each trial included applying 
one of the 0.25-templates or 0.75-templates to the cycle checked and 
calculating a $\chi^2$ test. Minimum $\chi^2$ has been considered 
indicative of the most plausible template allowing the turn-on time to be 
recovered. First we apply this procedure to 3 cycles with 0.5-orbital phase
turn-on time uncertainty, each time adding the newly categorized cycle
to the template, then the procedure was repeated for remaining 6 cycles.

The template points included mainly the main-on state and few points before
and after it. Note that the points near pre-eclipse X-ray absorption dips
(around orbital phase 0.75) have been excluded from templates because they
are strongly affected to the average X-ray intensity variations caused by
data gaps around dips. For the same reason, for some cycles with data gaps
preceding the turn-on, the first point with notable signal was excluded as
well.

The results of turn-on times determination are collected in Table 1. 
The first column of Table 1 contains the standard cycle number counted from
first UHURU observations. The second column shows to which 
intensity group, medium (m), high (h) or low (l), the cycle belongs.
The third column
displays modified Julian dates
($MJD=JD-2,400,000.5$). The fourth column contains the orbital phase
$\Phi_b$ of the turn-on . The fifth column shows
the cycle duration $\Delta$
expressed in binary orbits rounded to 0.5.
The sixth column shows the total number of points in the 
template used  (t) vs. the actual number of points of this template (u) 
in which $\chi^2$ test was calculated. The 
significance level is in 
the seventh column. For comparison, 
columns 8-9
are the same as 6-7, but for templates shifted by $-0.5$ cycles earlier 
the turn-on found.  Columns 10-11 are the same as 8-9 for
templates shifted by $+0.5$ cycles later the turn-on found.
Square brackets [] indicate that the template phase is rounded to 0.25 or
0.75.
For some cycles (e.g. \# 253, \#254, etc.)
we show the template minimization data for different number of template
points (the second row in columns 6-7).

We should shortly comment on Table 1. Practically for all cycles the
significance level is found to be $>70\%$. It is interesting to note that
although for cycle
\# 255 the turn-on time is well determined by the signal-to-noise ratio, the
use of a long template makes the probability of the cycle's turning-on half
an orbit later higher because the form of the X-ray main-on decrease of this
cycle is better fitted by 0.75-templates, but we preferred the turn-on time
determined by the real data. In addition, nearly for all remaining cycles
the probabilities of the turn-on $0.5$ orbit before/after the determined
value is very small, with the only exception of cycle \# 266, for which the
probability of turn-on half an orbit earlier is 87\%. However, this cycle is
poor in data.

As seen from Table 1, the number of 
cycles with a  duration of 20.5 orbits is notably larger than 
those of 20 or 21 orbits. Moreover, the first  nine successive 
cycles turned on after 20.5 orbits. Clearly, such a distribution 
is far from being random. There is a small correlation between 
the duration of a cycle and its mean X-ray flux, those of higher
intensity being longer (21 orbits) and of lower intensity shorter
(20 orbits). To increase the significance of this correlation 
a larger number of cycles should be analyzed.   

\begin{figure}
\epsfxsize=\columnwidth
\epsfbox{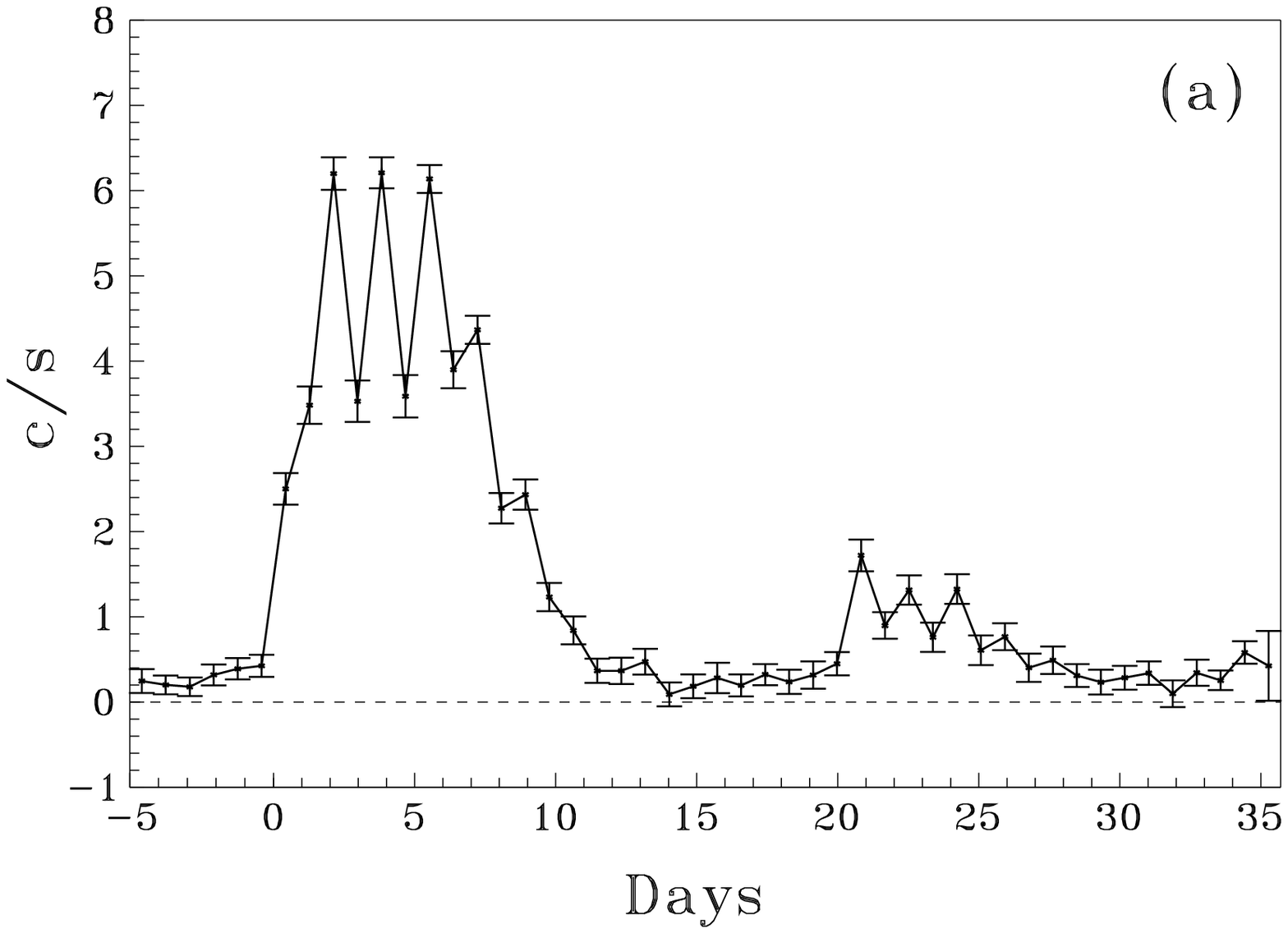} 
\epsfxsize=\columnwidth
\epsfbox{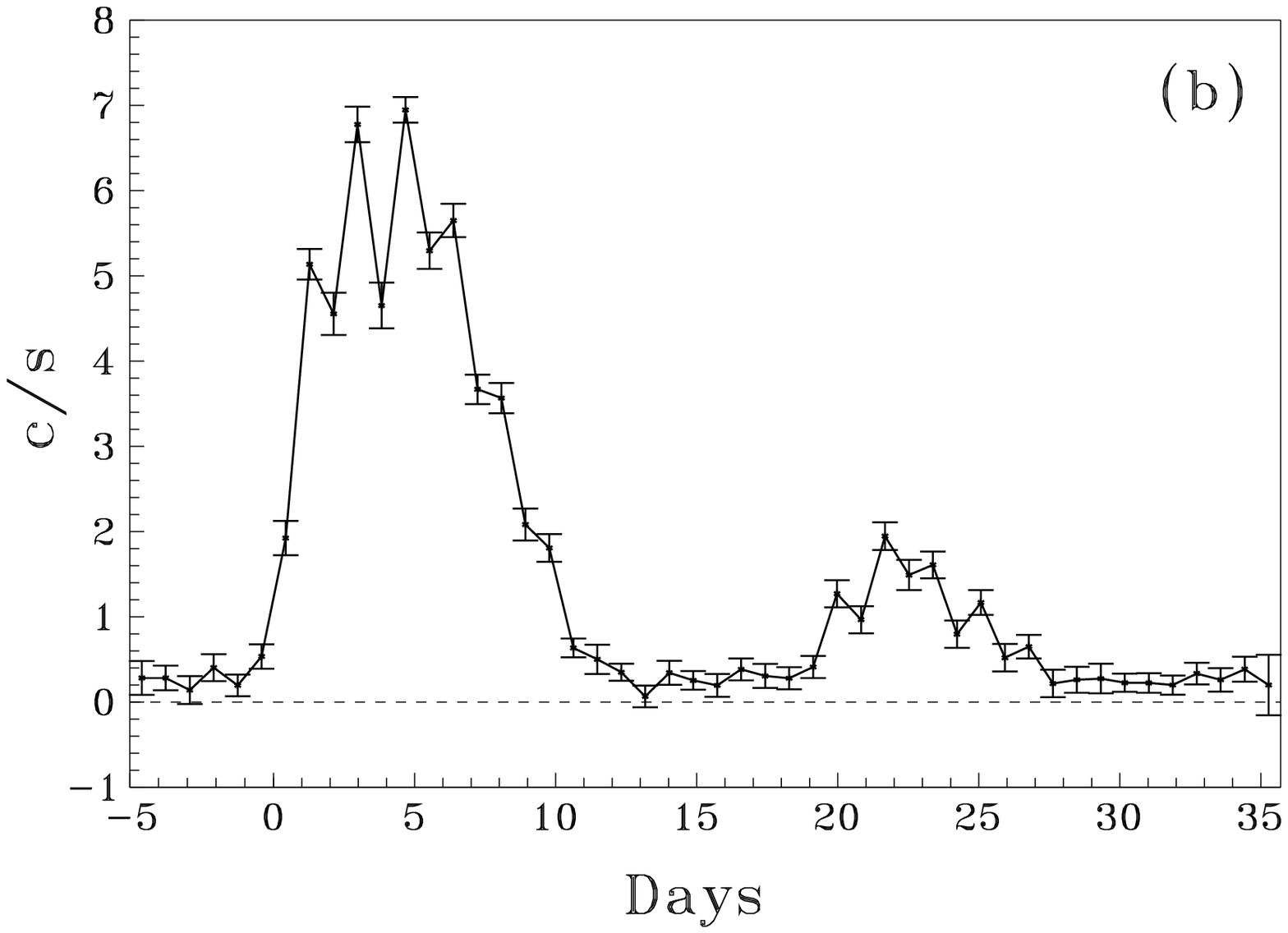} 
\caption{The mean RXTE X-ray light curve of Her X-1 for 
cycles turning-on at $[\Phi_b=0.25$ (a) and at $[\Phi_b=0.75$ (b)
obtained by folding the corresponding cycles without 
changing their actual duration.}
\end{figure}

\begin{figure}
\epsfxsize=\columnwidth
\epsfbox{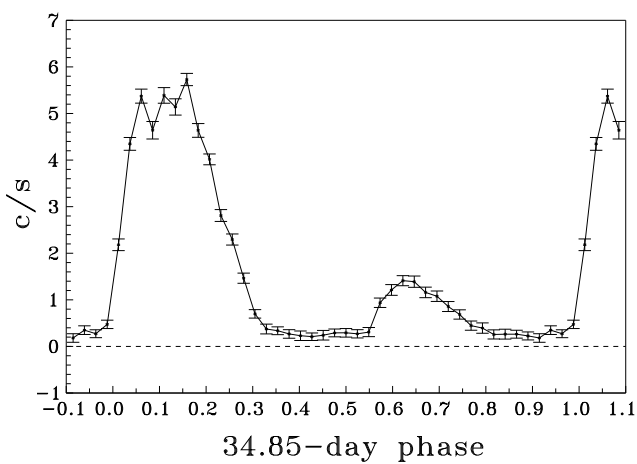} 
\caption{The mean X-ray light curve of Her X-1  obtained by adding 
all 0.25 and 0.75-cycles with stretching or compressing 
the cycles to a standard duration of  20.5 orbits.} 
\end{figure}

\subsection{The mean X-ray light curves}

Fig. 2(ab) display the mean X-ray light curves constructed by superimposing
separately 35-d cycles turned-on near orbital phase 0.25 (Fig. 2a) and
0.75 (Fig. 2b). The points inside the X-ray eclipses, i.e. in orbital phase
interval $0.932\div 0.068$, were excluded. In Fig. 2a (0.25-cycles), the
zero-point of 34.85-d period corresponds to the phase 0 of the orbital cycle
at which the turn-on occurred, while in Fig. 2b (0.75-cycles) it relates to
the phase 0.5 of the orbital cycle where the turn-on occurred. As seen from
Fig. 2ab, the pre-eclipse X-ray absorption dips are clearly visible on the
mean light curves. Moreover, they are also seen in the low-on state.

Fig. 3 shows the mean X-ray light curve obtained by adding the curves from
Fig. 2ab, assuming the average duration of each cycle to be exactly 20.5
orbits, for which we have stretched a little the cycles of shorter duration
and correspondingly compressed the longer ones. Due to zero-points
of the mean 0.25- and 0.75-light curves differing by $\Delta \Phi_b=0.5$
by construction, the absorption dips on the resulting total light curve
should have been compensated for. Nevertheless, the first pre-eclipse
dip on the 0.25-light curve is so strong that it is not compensated 
for and remains clearly visible on the mean X-ray light curve.     

In spite of a much smaller
statistics, the general shape of the mean RXTE X-ray light curve looks very
similar as found from UHURU data (Jones \& Forman 1976). Qualitatively, both
main-on and short-on stages are visible. During the main-on stage, the
source reaches the peak count rate in 1.5 orbits, stays at maximum for 2.5
orbits, and progressively fades to minimum during 3 orbits. The beginning of
the short-on stage is separated by 4.5 orbits from the end of the main-on.
The duration of the short-on state is about 5 orbits.

Note that the form of the short-on stage is similar to that of the main-on:
a rapid increase and slow decrease of count rate. The main-on state is
separated by a similar 4.5 orbits' off-state from the end of the short-on
state. A slow increase in count rate before the main-on state (first noted
by Jones \& Forman (1976)) is also clearly seen. A weak X-ray glow
persistent during off-states is possibly due to scattering in the accretion
disk corona.

\subsection{Dips and other features}

\begin{table}
\caption{Dips on the mean 0.25 and 0.75 X-ray light curves.}
\begin{center}
\begin{tabular}{clllcll}
\hline
\hline
Orb.  &  Orb.  & Error & Type &  Orb.  & Error & Type \\
cycle &  phase &       &      &  phase &       &      \\
\hline
\multicolumn{4}{c}{0.25}&\multicolumn{3}{c}{0.75}\\
\multicolumn{7}{c}{Main-on}\\    
\hline
    1   & 0.500  &  0.025  &   AI  & \dots   & \dots   &  \dots \\
        & 0.575  &  0.025  &   AM  & \dots   &  \dots  &  \dots  \\
        & 0.663  &  0.013  &   AE  & \dots   &  \dots  &  \dots  \\
        & 0.813  &  0.013  &   PI  & \dots   &  \dots  &  \dots  \\
        & \dots  & \dots   & \dots &  0.838  &  0.013  &   PI    \\
    2   & 0.788  &  0.013  &   PI  & \dots   &  \dots  &  \dots  \\ 
        & \dots  & \dots   & \dots &  0.813  &  0.013  &   PI    \\
    3   & 0.725  &  0.025  &   PI  & \dots  & \dots    & \dots \\
        & \dots  & \dots   & \dots &  0.788  &  0.013  &   PI    \\ 
        & 0.913  &  0.013  &   PBE &  0.913  &  0.013  &   PBE   \\
    4   & 0.713  &  0.013  &   PI  & \dots   & \dots   & \dots   \\
        & \dots  & \dots   & \dots &  0.738  &  0.013  &   PI    \\
        & 0.850  &  0.025  &   PBE & \dots   & \dots   & \dots   \\
        & \dots  & \dots   & \dots &  0.888  &  0.013  &   PBE   \\ 
    5   & 0.675  &  0.038  &   PI  & \dots   & \dots   & \dots   \\
        & \dots  & \dots   & \dots &  0.713  &  0.013  &   PI     \\
        & 0.825  &  0.038  &   PBE & \dots   & \dots   & \dots   \\
        & \dots  & \dots   & \dots &  0.850  &  0.025  &   PBE   \\
    6   & 0.650  &  0.038  &   PI  &  0.650  &  0.025  &   PI     \\
        & 0.750  &  0.038  &   PBE & \dots   & \dots   & \dots    \\
\hline
\multicolumn{7}{c}{Low-on}\\
\hline
    12 & 0.25:   &  \dots  &   TO & 0.25:   & \dots   &   TO \\
    13 & 0.17    &  0.025  &   RE & \dots   & \dots   &\dots \\
       & \dots   &  \dots  & \dots &0.20    &  0.05   &   RE \\
       & 0.45    &  0.05   &   AI & \dots   &  \dots  &\dots \\
       & \dots   &  \dots  & \dots &0.50    &  0.05   &   AI \\
       & 0.65    &  0.05   &   AE & 0.65    &  0.05   &   AE \\
       & 0.75    &  0.05   &   PI & 0.75    &  0.05   &   PI \\
    14 & 0.20    &  0.05   &   RE & 0.20    &  0.05   &   RE \\
       & 0.45    &  0.05   &   AI & \dots   & \dots   &\dots \\
       & 0.60    &  0.05   &   AE & \dots   & \dots   &\dots \\
       & 0.70    &  0.05   &   PI & 0.70    &  0.05   &   PI \\
    15 & 0.65    &  0.05   &   PI & 0.65    &  0.05   &   PI \\
    16 & 0.20:   & \dots   &   RE & \dots   & \dots   &\dots \\
       & 0.70:   & \dots   &   PI & \dots   & \dots   & \dots \\
       & 0.85:   & \dots   &   PE & \dots   & \dots   &\dots \\
    17 & 0.20:   & \dots   &   RE & \dots   & \dots   &\dots \\
\hline
\hline
\end{tabular}
\end{center}
{TO -- turn-on time;
AI, AM,  AE -- ingress to, minimum of, and egress from the anomalous dip;
PI, PBE -- ingress to and the beginning of egress from the pre-eclipse dip;
RE -- post-eclipse recovery.}
\end{table}

In order to investigate in more detail the form and features of
the RXTE X-ray light curve of Her X-1, 
we folded separately the cycles turning on at binary phases 0.25 and 0.75
and averaged the data over 20 equal bins in one orbital period (Fig. 4 and
5). Note that for these curves we used {\it all} available data, including
X-ray eclipses. The anomalous dip is clearly seen for 0.25-cycles during
the first orbit after the main turn-on and is practically invisible for
0.75-cycles. For both types of cycles, the pre-eclipsing dips demonstrate the
identical behaviour -- they march from the eclipse toward earlier orbital
phase in successive orbits. In Table 2 we list the orbital phases of ingress
to and egress from the pre-eclipse dips (PI, PE) and the anomalous dip (AI,
AE). Note also that post-eclipse recovery (RE) is not present in the mean light
curves. 

Both for 0.25 and 0.75-cycles, the low-on state turns on in the 12th orbit
after the main turn-on. After the turn-on,  both post-recovery and anomalous 
dips are present during two successive orbits. In the 15th orbit, the X-ray
source emerges immediately after the eclipse and the anomalous dip is
absent. In the 16th and 17th cycles, a post-recovery dip may be 
marginally distinguished for 0.25-cycles. The pre-eclipse dips 
march likely to the main-on state.

\begin{figure*}
\fbox{\vbox{
\centerline{}
\centerline{This figure is avaliable at URLs:}
\centerline{}
\centerline{\bf http://xray.sai.msu.su/pub/preprints/Prokhorov/rxte/fig4a.ps}
\centerline{or}
\centerline{\bf http://xray.sai.msu.ru/pub/preprints/Prokhorov/rxte/fig4a.ps}
\centerline{}
}}
\end{figure*}

\begin{figure*}
\fbox{\vbox{
\centerline{}
\centerline{This figure is avaliable at URLs:}
\centerline{}
\centerline{\bf http://xray.sai.msu.su/pub/preprints/Prokhorov/rxte/fig4b.ps}
\centerline{or}
\centerline{\bf http://xray.sai.msu.ru/pub/preprints/Prokhorov/rxte/fig4b.ps}
\centerline{}
}}
\caption{The mean RXTE light curve of Her X-1 for all 34.85-d cycles
that turn on near $\Phi_b=0.25$. The vertical quadrangles mark X-ray eclipses.
Points are individual RXTE observations. 
These points are averaged in 
20 bins per one orbit. 
Bars show the mean count rate plus/minus rms error
inside these bins.}
\end{figure*}

\begin{figure*}
\fbox{\vbox{
\centerline{}
\centerline{This figure is avaliable at URLs:}
\centerline{}
\centerline{\bf http://xray.sai.msu.su/pub/preprints/Prokhorov/rxte/fig5a.ps}
\centerline{or}
\centerline{\bf http://xray.sai.msu.ru/pub/preprints/Prokhorov/rxte/fig5a.ps}
\centerline{}
}}
\end{figure*}

\begin{figure*}
\fbox{\vbox{
\centerline{}
\centerline{This figure is avaliable at URLs:}
\centerline{}
\centerline{\bf http://xray.sai.msu.su/pub/preprints/Prokhorov/rxte/fig5b.ps}
\centerline{or}
\centerline{\bf http://xray.sai.msu.ru/pub/preprints/Prokhorov/rxte/fig5b.ps}
\centerline{}
}}
\caption{The same as in Fig. 4 for the cycles turning on near $\Phi_b=0.75$.}
\end{figure*}

\section{Accretion disk model constraints}

The durations of main-on and short-on states may be used to put
bounds on disk geometrical parameters: its inclination to the orbital plane
$\delta$  and
the semi-thickness-to-radius ratio  $H/R$. The disk is assumed to be tilted
with respect to the orbital plane and to counter-precess due to tidal
interaction with the optical star.

\begin{figure}
\epsfxsize=\columnwidth
\epsfbox{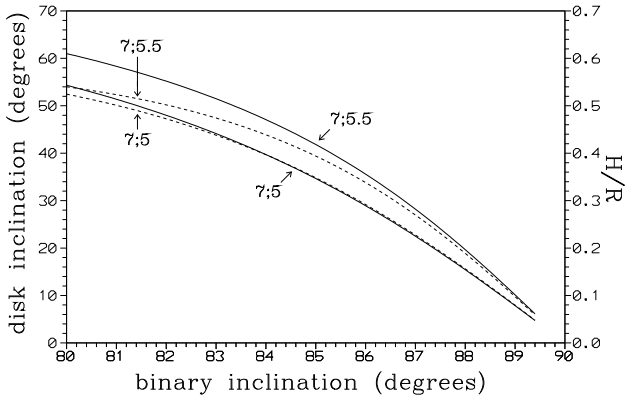} 
\caption{The disk inclination angle $\delta$ (solid lines) and its
dimensionless semi-thickness $H/R$ (dashed lines) as a function of
binary inclination angle $i$. The lines are labeled with the
duration of the main-on (7) and short-on (5 or 5.5) states in orbital periods.}
\label{fig6}
\end{figure}

From the duration of the main-on $d_m$ and low-on 
$d_l$ states we can determine the dependence of 
the disk  
inclination to the orbital plane $\delta$ and the parameter 
$H/R$, on the binary inclination $i$.
The angle between the disk normal and the line of sight  
$\theta$ relates with
the azimuthal angle of disk precession $\phi$ as
$$
\cos \theta = \cos\delta \cos i + \sin\delta \sin i \cos\phi \,.
$$
For the main-on state $\theta=\theta_m \equiv \pi-\arctan (H/R)$,
whereas for the low-on state $\theta=\theta_l \equiv -\theta_m$.
Using these relations we obtain
$$
	\delta = - \arctan \left( \frac{2}{\tan i (z_m+z_l)} \right)
$$
and
$$
	H/R = \arcsin \left( \frac{z_m-z_l}{2} \sin\delta \sin i \right) \,,
$$
where
$$
	z_m \equiv \cos(2\pi d_m / P_{prec}) \,,\qquad
	z_l\equiv \cos(2\pi d_s / P_{prec}) \,,
$$
and $P_{prec}$ is the disk precession period. 
Fig. \ref{fig6} shows $\delta$
and $H/R$ as a function of the orbital inclination $i$ 
for limiting durations $d_m$ and $d_l$ expressed in orbital cycles.

\section{Conclusions}

The analysis of the publically available RXTE data on Her X-1 confirmed main
features of 34.85-d cycle of Her X-1. We have determined turn-on
times of new RXTE cycles with a good accuracy. The use of template
light curves has enabled us to find turn-on times of the cycles
with data gaps near the turn-on. We discovered that the number of 
cycles with a duration of 20.5 orbits is significantly higher
than those of shorter or longer duration, thus casting 
doubts as to the randomness in the successive cycle durations.
Instead, a small correlation between the mean X-ray flux and the cycle
duration has been found. Such a correlation has also been noted in 
BATSE data (Wilson et al. 1994). This correlation is possibly related to 
the dynamical action of accreting streams on the precessional motion of 
the accretion disk, with the tidal interaction forcing the disk to
counter-precess while the dynamical action pushing the disk into 
co-orbital precession. The higher accretion rate, the stronger 
action of streams, hence, the longer the net precessional cycle 
duration is. 

Using the mean X-ray light curve,
we have found that the form of the secondary turn-on state looks similar to
that of the main-on state, i.e. is characterized by a rapid increase and a slow
decrease, with the total duration of 5-5.5 orbital cycles.
We also studied the absorption features (dips) and post-recovery on the mean
X-ray light curve. We have found that during the main-on state
the anomalous dip is present only in one orbit after the turn-on, whereas
during low-on state, the anomalous dip is clearly visible during
two orbits after the turn-on. The post-eclipse recovery have not been
found in main-on state but appears at least two times during low-on
state. The pre-eclipsing dips are found in both main-on and low-on states
and demonstrate the behaviour as in previous observations.

\section*{Acknowledgments}
The authors acknowledge the referee, Prof. F.Nagase, for pointing to an
inaccuracy in our analysis, and for suggesting the use of templates in the
determination of the turn-on times for cycles with data gaps. Research has
made use of data obtained through the High Energy Astrophysics Science
Archive Research Center Online Service, provided by the NASA/Goddard Space
Flight Center. This work is partially supported by the RFBR grant
98-02-16801, by the NTP program "Astronomija" (project 1.4.4.1) of Ministry
of Science and High Technology, and by INTAS grant 93-3364-Ext.



\begin{thebibliography}{99}

\bibitem{} Bahcall J.N., Bahcall N.A., 1972, ApJ, 178, L1

\bibitem{} Baykal A., Boynton P.E., Deeter J.E., Scott D.M., 1993, MNRAS,
265, 347

\bibitem{} Bildsten L. et al., 1997, ApJSS 113, 367

\bibitem{} Brecher K., 1972, Nat, 239, 325

\bibitem{} Cherepashchuk A.M., Efremov Yu. N., Kurochkin N.E., Shakura N.I.,
Sunyaev R.A., 1972, Inform. Bull. Variable Stars 720

\bibitem{} Cooke B.A., Page C.G., 1975, Nat, 256, 712

\bibitem{} Crosa L.M., Boynton P.E., 1980, ApJ, 235, 999

\bibitem{} Davison P.J.N., Fabian A.C., 1974, MNRAS, 169, 27P

\bibitem{} Davison P.J.N., Fabian A.C., 1977, MNRAS, 178, 1P

\bibitem{} Deeter J.E., Boynton P.E., Pravdo S.H., 1981, ApJ, 247, 1003

\bibitem{} Deeter J.E., Boynton P.E., Miyamoto S., Kitamoto S., Nagase F.,
Kawai N., 1991, ApJ, 383, 324

\bibitem{} Fabian A.C., Pringle J.E., Rees M.J., 1973, Nat, 224, 212

\bibitem{} Gerend D., Boynton P.E., 1976, ApJ, 209, 562

\bibitem{} Giacconi R., Gursky H., Kellogg E., Levinson R., Schreier E.,
Tananbaum H., 1973, ApJ, 184, 227

\bibitem{} Gorecki A. et al., 1982, ApJ, 256, 234

\bibitem{} Howarth I.D., Wilson B., 1983, MNRAS, 202, 347

\bibitem{} Jones C.A., Forman W.H., Liller W., 1973, ApJ, 182, L109


\bibitem{} Jones C., Forman W., 1976, ApJ, 209, L131

\bibitem{} Levine A.M., Jernigan J.G., 1976, ApJ, 262, 294

\bibitem{} Liller W., 1972, IAUC, 2415

\bibitem{} Kurochkin N.E., 1972, Peremennye Zvezdy, 18, 425

\bibitem{} Mironov A.V., Moshkalev V.G., Trun\-kov\-skij E.M.,
Che\-re\-pa\-shchuk A.M., 1984, PAZh, 10, 429

\bibitem{} Nagase F., Hayakawa S., Kii T. et al., 1984, PASJ, 36, 667 

\bibitem{} Novikov, I.D., 1973, AZh, 50, 459

\bibitem{} \"Ogelman H., Kahabka P., Pietsch W., Tr\"umper J., Voges W.,
1985, Space Sci. Rev., 40, 347

\bibitem{} \"Ogelman H., 1987, A\&A, 172, 79

\bibitem{} \"Ogelman H., Tr\"umper J., 1988, in Pallavicini R., White N.E.,
eds., X-ray Astronomy with EXOSAT, Mem. S.A.It., 59, 169

\bibitem{} Ohashi D., Inoue H., Kawai N., Koyama K., Matsuoka M., Mitani K.,
Tanaka Y., 1984, PASJ, 36, 719

\bibitem{} Ricketts M.J., Stanger V., Page C.G., 1982, in Brinkmann W.,
Tr\"umper J., eds, Accreting Neutron Stars, ISO (Garching bei M\"unchen), 100

\bibitem{} Shakura N.I., Postnov K.A., Prokhorov M.E., 1998, A\&A, 331, L37

\bibitem{} Sheffer E.K., Kopaeva I.F., Averintsev M.B. et al., 1992, AZh,
69, 82

\bibitem{} Shklovskii I.S., 1973, AZh, 50, 233

\bibitem{} Soong Y., Gruber D.E., Rothschild R.E., 1987, ApJ, 319, L77

\bibitem{} Staubert R., Bezler M., Kendziorra E., 1983, A\&A, 117, 215

\bibitem{} Tananbaum H., Gursky H., Kellogg E.M., Levinson R., Schreier E.,
Giacconi R., 1972, ApJ, 174, L143

\bibitem{} Tr\"umper J., Kahabka P., \"Ogelman H., Pietsch E., Voges W.,
1986, ApJ, 300, L63

\bibitem{} Von Wentzel W., Gessner H., 1972, Mitt. Ver. Sterne, 6, 61

\bibitem{}  Wilson R.B., Finger M.H.,
Pendelton G.N., Briggs M., Bildsten L., 1994, in  Holt S.S.,
Day C.S., eds,  The Evolution of X-ray
Binaries,  AIP Press (New York), 475

\end{thebibliography}
\end{document}